\documentclass[a4paper,conference]{IEEEtran} 
\ifCLASSINFOpdf
  \usepackage[pdftex,dvipdfm]{graphicx}
\else
\fi
\usepackage[cmex10]{amsmath}
\usepackage{amssymb,verbatim}
\usepackage[dvipdfm]{graphicx}
\begin{document}
%
\title{%
Digest of Quantum Stream Cipher based on Holevo-Yuen Theory
}

\author{
\IEEEauthorblockN{Masaki SOHMA$^{1}$, Osamu HIROTA$^{1,2}$ \\}
\IEEEauthorblockA{
1. Quantum ICT Research Institute, Tamagawa University\\
6-1-1, Tamagawa-gakuen, Machida, Tokyo 194-8610, Japan\\
2. Research and Development Initiative, Chuo University, \\
1-13-27, Kasuga, Bunkyou-ku, Tokyo 112-8551, Japan\\
{\footnotesize\tt E-mail:sohma@eng.tamagawa.ac.jp,  hirota@lab.tamagawa.ac.jp} \vspace*{-2.64ex}}
}

\maketitle

\begin{abstract}
So far, quantum key distribution (QKD) has been the main subject in the field of quantum cryptography, but that is not quantum
 cryptographic communication, it is only the ability to send keys for cryptographic purposes. 
To complete cryptographic communication, a technique for encrypting data is necessary, and the conventional cryptographic
 technique of mathematical symmetric key cipher  or  One Time Pad (OTP)  is adopted in the discussion so far. 
However, OTP is not the ultimate cipher for data encryption, because it does not satisfy security conditions in the modern cryptology. 
 Around 2000, a new quantum stream cipher was proposed as a technique to challenge the possibility of overcoming  drawbacks
 of OTP  in practical use. Recently, we have published some review papers on it in Entropy (Open access journal) [1], and others [2,3].
This paper introduces an overview and a back ground of our paper that is entitled Quantum stream cipher based on Holevo-Yuen theory.
\end{abstract}

%
\IEEEpeerreviewmaketitle

\section{General View of Cryptography or Cipher in Social Network Systems }

Around 2000, the  government and communication service providers  have imposed the conditions shown in Fig. 1 on the development for future telecommunications security technology. 
Then Y-00 quantum stream cipher was proposed as a cryptographic technique that satisfies these conditions and is currently undergoing commercialization. 

In the recent Book [4] and a technical paper [5], S. Tsujii who is one of the leaders of the cyber security community 
and industry  explains the current situation of the cyber security community 
and industry on the  trend of the security technology as follows. \\
``Quantum computer capable of breaking public key cryptographies,  such as RSA  or elliptic curve cryptography,  
that relies on mathematical decipherability due to prime number factorization or discrete logarithm problems, will not be developed within 20 years.
 Nevertheless, the jeopardy due to the cooperative effect with the development of mathematics remains. 
Thus, NIST is in the process of selecting candidates for quantum computer-resistant cryptography (see Appendix [A]). The applications of cryptography
 for confidentiality are categorized into the confidential transmission of data itself and the key delivery or storage for that purpose. 
Then from the viewpoint of academic methods, they are categorized into mathematical cryptography and quantum cryptography. 
In the former case, there are two types  such as public key cryptography and symmetric key cipher. 
Public key cryptography has the advantage of securely delivering and storing the initial key for data encryption and transmission. 
But its processing speed is slow,  so symmetric key cipher is responsible for data encryption.
On the other hand, quantum cryptography is a cryptographic technique that uses quantum phenomena to improve security performance. 
The technique that uses quantum communication to perform the key delivery function of public key cryptography is 
quantum key distribution (QKD: BB-84 et al), while the technique that uses quantum communication to perform the cryptographic transmission of data itself
 is called Y-00 quantum stream cipher (see Fig. 2). QKD cannot be used to supply keys to One Time Pad cipher, because its data rate is too slow. 
Y-00 for data encryption is extremely novel in its ability to prevent eavesdroppers from obtaining the ciphertext of the symmetric key cipher. 
 In addition, it is amazing that the strong quantum-ness is created by modulation scheme with multi-ary coherent state signals without any quantum device."

\begin{figure}
\centering{\includegraphics[width=8cm]{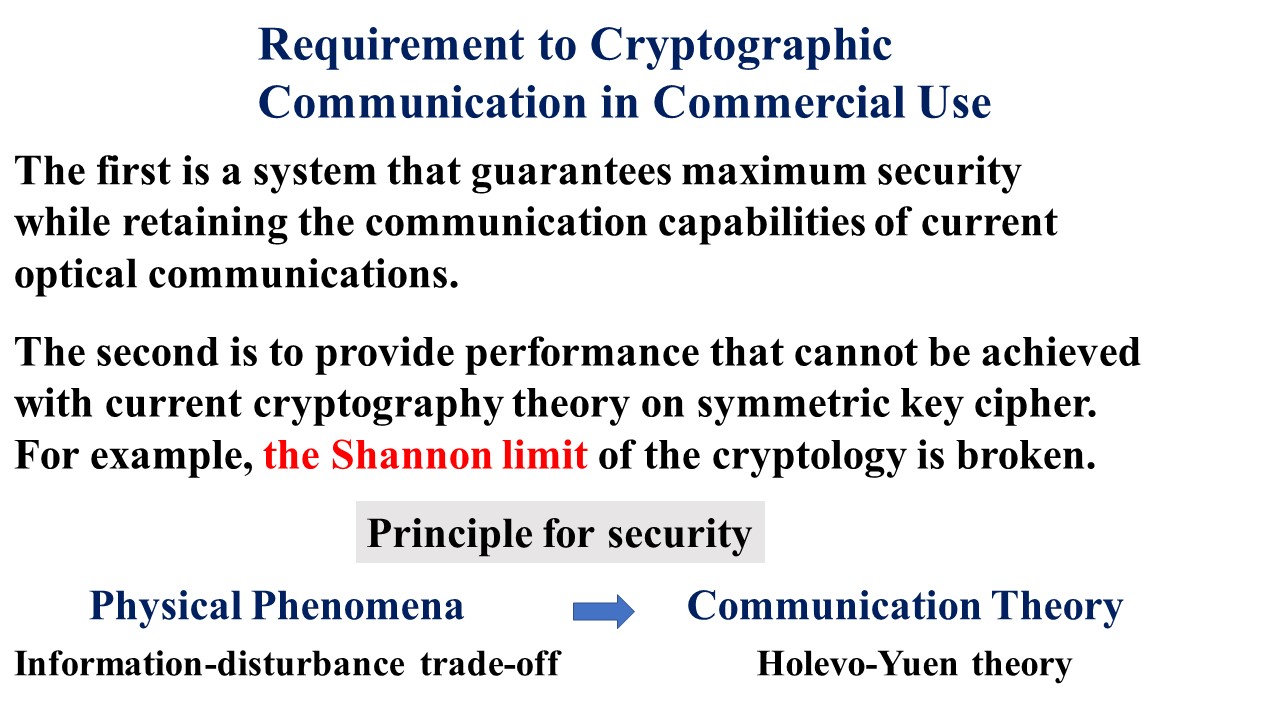}}
\caption{Basic requirements of performance to new technologies}
\end{figure}

\begin{figure}
\centering{\includegraphics[width=8cm]{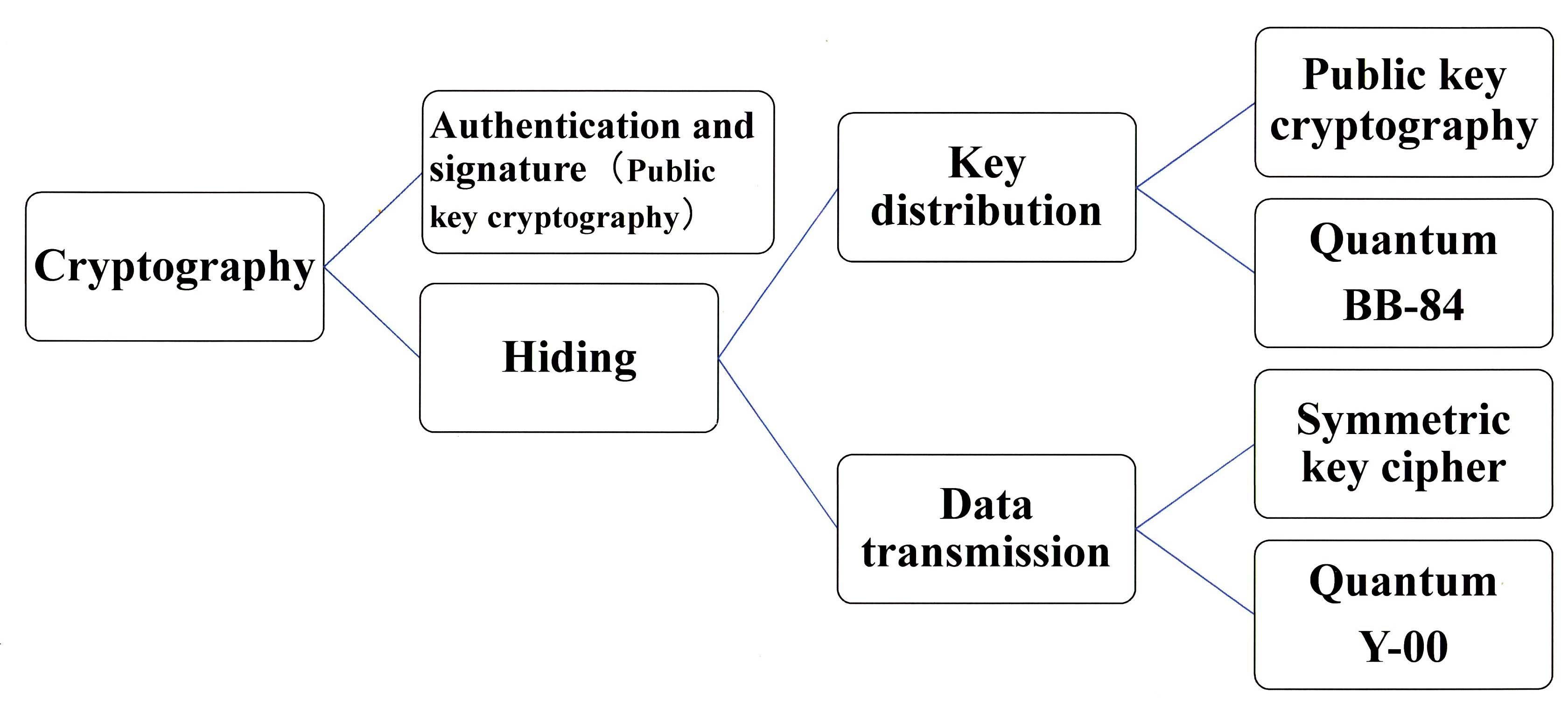}}
\caption{Classification of cryptographic techniques}
\end{figure}

Let's now turn our focus to quantum cryptography.
Both of these quantum technologies are based on designing communication systems to make it difficult for eavesdroppers to steal signals on
 the communication channels. Such a function to protect the signal itself cannot be realized by mathematical cryptography. 
As mentioned above, there are two possible system operation methods for these quantum cryptography techniques.  
One is to use BB-84 quantum key distribution for key delivery and conventional mathematical cryptography for authentication and data encryption. 
The other is to use Y-00 quantum stream cipher for data encryption and conventional public key cryptography (or quantum computer resistant type) 
for authentication and key delivery.
These quantum cryptography technologies are positioned as technologies to ensure the ultimate security of communication between data center stations, 
that is of special importance in next-generation 5G and 6G systems.
 In the following, we will explain the technical contents, applicability to the real world, and development trends.

\section{ Current Status of Quantum Communication Security Technology}
\subsection{Quantum cryptography}
As introduced in the above section, there are two quantum cryptography techniques. Let us give their brief introduction  below.\\

(1) Quantum Key Distribution\\

BB-84 quantum key distribution (QKD) was proposed by C. H. Bennett and G. Brassard in 1984. It is a protocol to share a secret key sequence by using photon communication,
 that is guaranteed to be quantum  nature. Since the photons used in this protocol are weak light, the transmission speed and distance are limited.
 In addition, many of the sequence of photons that carry information are lost due to attenuation effects in the transmission line,
 and the sequence of photons that reaches the receiver is also subject to errors due to noise effects.
So the operation involves discarding the majority of the received bit sequence. Therefore, data itself cannot be sent, only random numbers can be sent. 
Thus only the delivery of the secret key for symmetric key cipher is possible. This is why it is called QKD. 
Recently, many newspapers have reported that several R $\&$ D groups can provide the commercial systems of QKD. 
The transmission speed is the order of 100 Kbit/sec, and transmission length is below 100 Km.
 The satellite system is one of the solutions to cope with the distance. But the transmission speed is so small.
In any case, if one tries to increase the transmission speed, there is a trade-off  and one has to shorten the relay interval. 
Since the maximum transmission speed is about a megabit, it is difficult to supply keys to the One Time Pad cipher for data after key delivery,
 and it is likely to be limited to supplying initial keys (secret keys) for  AES and others (See Appendix [B]).\\

(2) Quantum Stream Cipher\\

Y-00 quantum stream cipher is a protocol for physical symmetric key cipher proposed by H.P. Yuen of Northwestern University in the DARPA project (2000)  [6]. 
The details are explained in the next section, but a simple concept is presented here. 

\begin{figure}
\centering{\includegraphics[width=8cm]{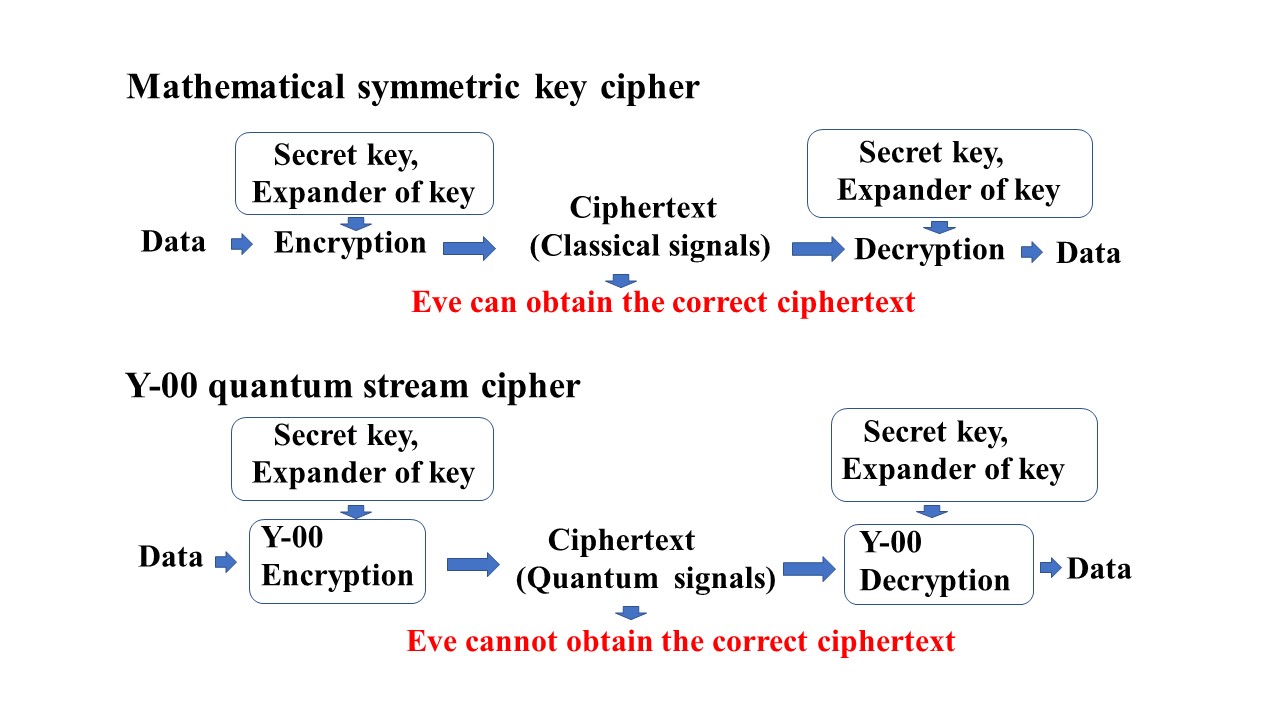}}
\caption{Principle of operation of Y-00 quantum stream cipher. The expander of key in the both cases means PRNG
  that is employed in  the conventional cipher system.  Classical signal means that they have distinguishability, and quantum signal means impossible
 to distinguish them precisely. Y-00 Encryption is the function of converting a classical signal into a quantum signal. It is also called quantum modulation.
 }
\end{figure}

This technique is characterized by the fact that it does not allow the physical signals consisting of 
the mathematical ramdom generator and information data to be obtained without error.
 In this scheme, the ciphertext in Y-00 circuit system of the mathematical cipher consisiting of the generator and data, which is the target of the eavesdropper, 
is described by $ y=\alpha_i(X,f_g(K_s),R_p) $.
Then, we design the system such that the ciphertext  $ y=\alpha_i(X,f_g(K_s),R_p) $  is mapped into ensemble of coherent state $\mid \Psi (X,K_s, R_p)>$
with the  quantumness  based on Holevo--Yuen theory [7,8,9].
This is called Y-00 signal which corresponds to ciphertext on the Hilbert space.
Thus, the ciphertext as the classical signal is protected by the quantumness. Let us describe it shortly. 
Although ordinary laser light of high power is used as the transmission signal, signals on the communication channel can be made to
 have very strong quantum properties in the sense of quantum detection theory. This is Y-00 principle [6]. 
That is,  a large number of physical binary light communication base is prepared to transmit electric binary data, and the binary data is transmitted by using
 one communication base which is randomly selected from many communication bases by a mathematical cipher. Let $M$ be the number of the base. 
 The optical signals on the communication channel become ultra-multiple-valued signals ($2M=4096$ or more values are common)  against
 the eavesdropper without the knowledge of communication base. 
At this time, strong quantum nature in the signal ensemble appears even if  the one signal is in high power light, when it is constructed by such ultra-multiple-valued. 
In other words, this method means that the quantum nature in the sense of quantum detection theory is created artificially by modulation schemes, 
so that it does not require light with strong physical quantum nature such as photon. The Y-00 signals of the length $m$ (number of slot) are described as follows:
\begin{eqnarray}
&&\mid \Psi (X,K_s, R_p)> =\mid \alpha_i(X,f_g(K_s),R_p) >_1  \nonumber \\
&&\otimes \mid \alpha_j(X,f_g(K_s),R_p) >_2  \dots  \dots \nonumber \\
&&\otimes \mid \alpha_k(X,f_g(K_s),R_p) >_m
\end{eqnarray}
where $\mid \alpha_i(X,f_g(K_s),R_p) >$ is coherent state with amplitude $\alpha(\cdot)$, $i,j,k =1,2,3, \dots 2M$, $X$ is plaintext, $f_g(K_s)$ is
 a mathematical pseudo random function of  secret key $K_s$, and $R_p$ is additional randomization. The set of these coherent states 
is designed to be strong non-orthogonal property, 
even if each amplitude of the signals is  $ |\alpha_k(X,f_g(K_s),R_p) | \gg 1$.

A legitimate receiver with the knowledge for communication base to which the data is sent can ignore the quantum nature of the data, because it is a binary
 transmission by high power signal. That is, he can receive the data of error-free. 
On the other hand, an eavesdropper, who does not know 
the information of communication base, must receive a sequence of a ultra-multi-valued optical signal that consists of non-orthogonal quantum states of Eq(1).
The quantum noise generated by quantum measurement based on Holevo-Yuen theory on quantum detection masks the received signal, 
resulting in errors.
 Thus, even if the eavesdropper tries to record the ciphertext, the masking effect of the quantum noise makes it impossible to accurately recover the ciphertext. 
This fact is a novel function in the cryptology.  Fig 3 shows the scheme of Y-00 protocol. And Fig 4 shows the  experimental demonstration of the advantage creation principle of security based on Holevo-Yuen theory.

\begin{figure}
\centering{\includegraphics[width=8cm]{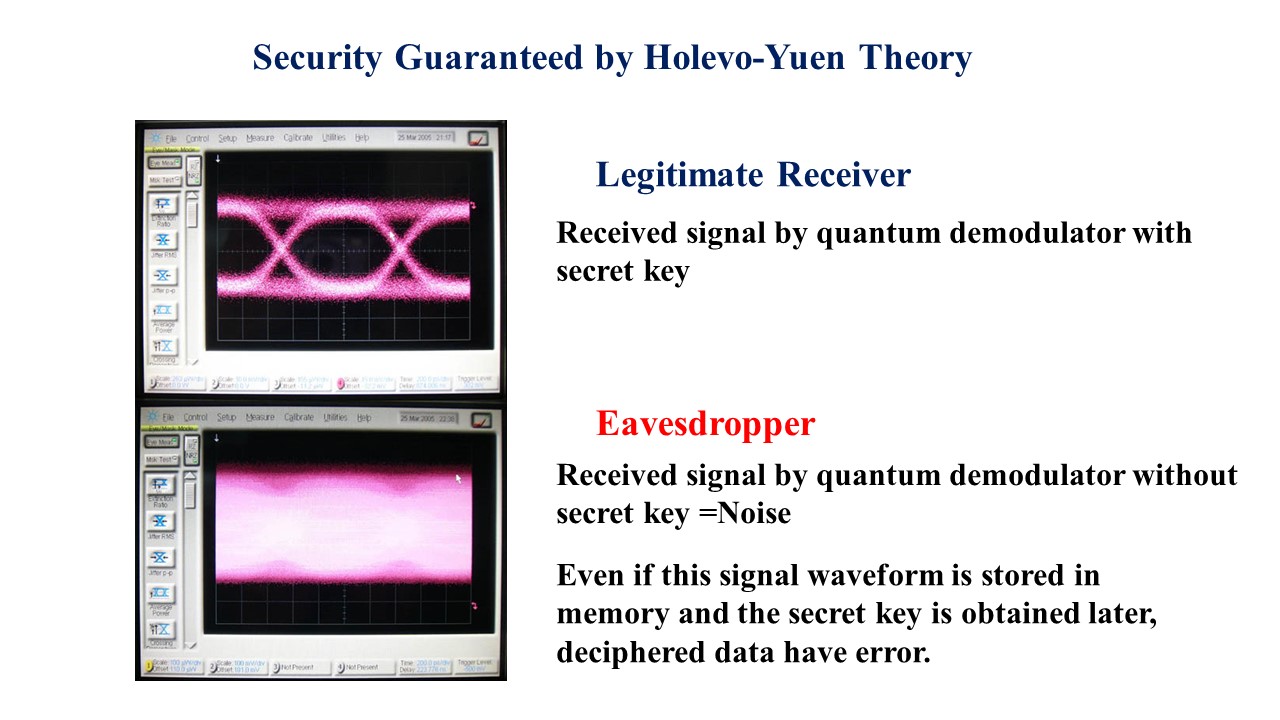}}
\caption{Experimental demonstration of advantage creation based on Holevo-Yuen theory. Quantum ciphertext  for eavesdropper consists of 2$M$ densely packed non-orthogonal quantum coherent state signals. As a result, Holevo-Yuen theory guarantees that an eavesdropper cannot receive the correct ciphertext, or cannot copy the ciphertext.}
\end{figure}

\subsection{Comparison of services based on each quantum cryptosystem}
QKD and Y-00 are about 40 and 20 years old, respectively, since they were invented. 
At the time of their invention, the principle models of both quantum cryptography technologies were not very attractive in terms of security
 and communication performance.
But nowadays, the systems and security assurance technologies of both technologies have evolved dramatically. 
Based on the results, business models for security services using these quantum cryptography technologies have been proposed.
 Fig.5 shows  the cryptography communication scheme based on two types of quantum cryptographies, 
and Fig.6 shows the current status of the system performance.

\begin{figure}
\centering{\includegraphics[width=8cm]{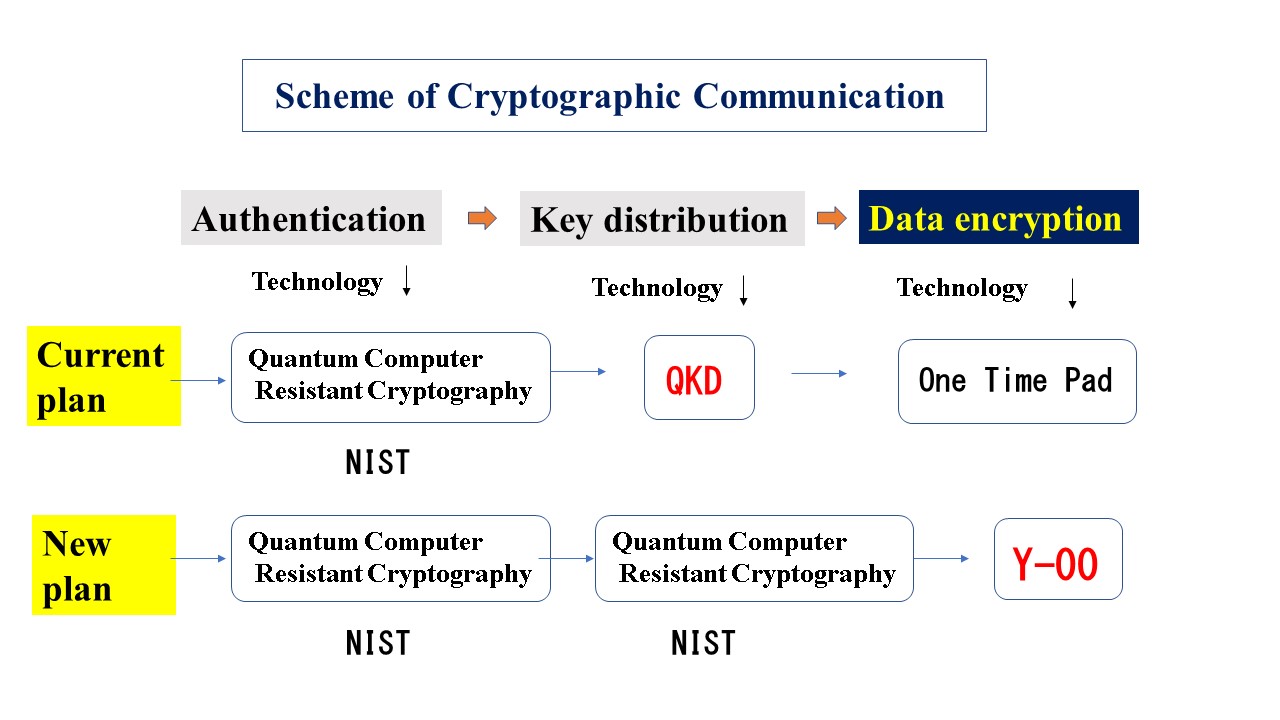}}
\caption{Two types of quantum cryptographic communication schemes }
\end{figure}

\begin{figure}
\centering{\includegraphics[width=8cm]{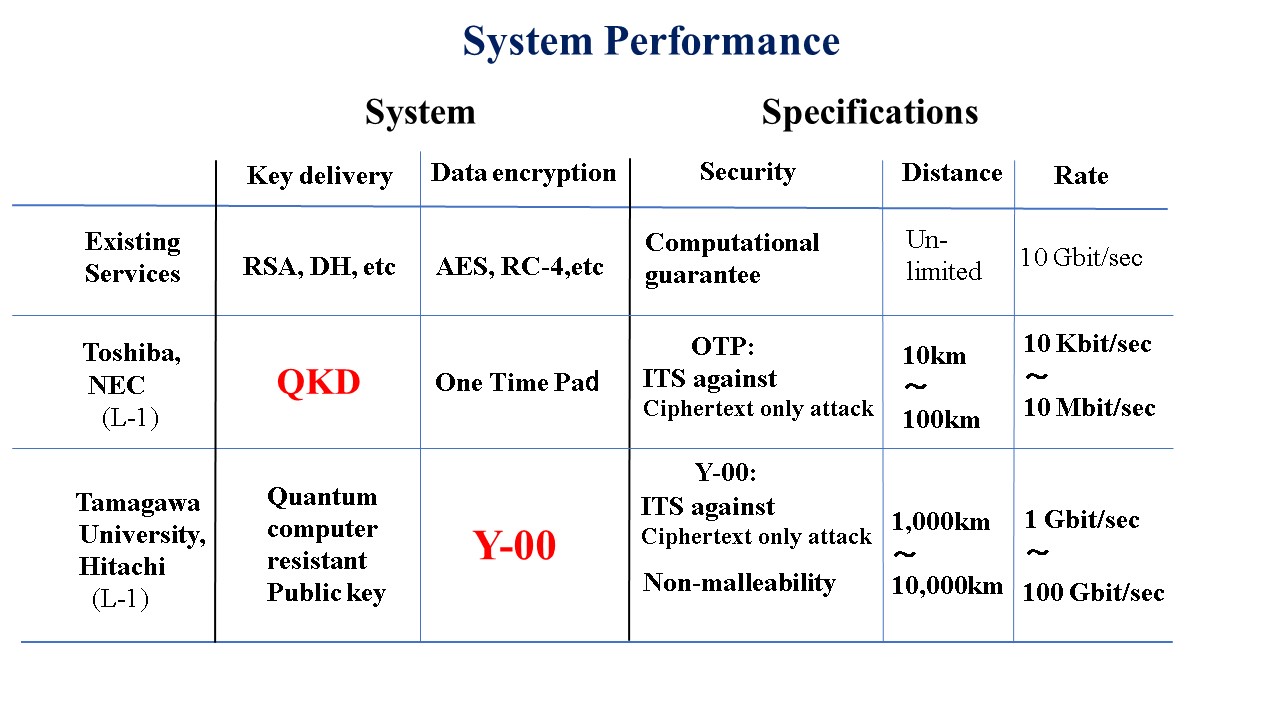}}
\caption{Comparison of product capabilities for two types of quantum cryptography services }
\end{figure}

\section{Feature of quantum stream cipher }
In the near future, optical networks will move toward even higher speeds, but Y-00 quantum stream cipher can solve technical requirement from the real world. 
Since there are few introductions to this technology, we describe the details of this technology at the following.

\subsection{Basic Scheme }
As explained in the previous section, the quantum stream cipher is expected to accelerate advanced application in the future communication system. 
The reason for this is that this scheme can utilize ordinary optical communication devices and is compatible with existing communication systems.
 In its design, optical communication, quantum theory, and cryptography are effectively integrated. 
Therefore, it is also called "Y-00 optical communication quantum cryptography" in implementation studies. 
Pioneering researches on practical experiment for this system have been reported by Northwestern University [10,11], Tamagawa University-Panasonic [12], and Hitachi Ltd [13].  
Theories of system design for the basic system have been given by Nair and others  [14,15,16,17].

Let us explain the principle of Y-00 quantum stream cipher. First, Y-00 protocol starts by specifying the signal system that use as the transmission medium. 
The actual signal to be transmitted is selected in terms of amplitude or intensity, phase, quadrature amplitude, etc., having coherent state $|\alpha \rangle $ in quantum optics. 
Then the design is made accordingly. 
Depending on the type of signal to be used, it is called as ISK:Y-00, PSK:Y-00, QAM:Y-00, etc. 

Here, one communication base consisting of  various binary signals is randomly selected by PRNG (or AES) in each data slot. Then a binary data is transmitted by using the communication
 base selected. 
Thus ultra-multi-valued  signals appear to be transmitted on the channel. The eavesdropper has to receive the ultra-multi-valued signal, because she does not 
which communication base was selected. 

\subsection{Progress in Security Theory}
The BB-84 protocol is a key delivery technique for securely sharing secret key sequences (random numbers).
 The Y-00 protocol is a symmetric key stream cipher technique for cryptographically transmitting data. 
As mentioned above, both quantum cryptography techniques enhance security by preventing eavesdroppers from taking the exact signal on the communication channel. 
The models that explains the principle of such physical technology are called the ``basic model". It is this basic model that can be found in textbooks for beginners. 

Let us start with QKD such as BB-84.
If the basic model of the BB-84 protocol is implemented in a real optical fiber communication system, it can be eavesdropped. 
Therefore, in order to guarantee security even in systems with noise and energy loss, a technique that combines  error correction and privacy amplification
 (universal hashing) was proposed, and then  a theoretical discussion of security assurance became possible. 
That is, in 2000, P. Shor, et al  proposed a mathematical security  theory for BB-84 on an abstract mathematical model called the Shor model, 
which was later improved by R. Renner.
 In brief, the security of the BB-84 protocol is evaluated by quantifying quantum trace distance of the two density operators  to the ideal random sequence
 and the random sequence shared by the real system. 
This is the current standard theory for the security of QKD. It is very difficult to realize a real system that 
 the quantum trace distance is sufficiently small.

On the other hand, from the beginning, Y-00 protocol can consider the effects of non-ideal communication systems. 
As mentioned at the above section, the selection of communication base of Y-00 protocol is encrypted by conventional mathematical cipher.
Y-00 quantum ciphertext, which is an optical signal, is emitted as the transmission signal.
So, the ciphertext of the mathematical symmetric key cipher that an eavesdropper needs to decipher corresponds to Y-00 quantum ciphertext.
However, since the set of ultra-multi-valued signals, which is  Y-00 quantum ciphertext, are non-orthogonal quantum state ensemble, 
her received signals are inaccurate due to errors caused by quantum noise. 
Therefore, the discussion based on the computational security of the mathematical cryptographic part of  Y-00 mechanism to be attacked is 
replaced by the problem of combination of information theoretic analysis and computational analysis. 
However, we should emphasize that the discussion with infinite  number or asymptotic theory are not our concern, because our concern is a physical system 
under  practical situation. For example, if attacker needs circuits of number of the size of the universe to perform the brute-force attack, the system is unbreakable.
Or, if attacker needs 100 years to collect the ciphertext for trying the crypto-analysis, it is also unbreakable.

\subsection{Randomization technology for quantitative security performance (\textbf {Errata of the original paper [1]})}
In the early days when Y-00 was invented, the model was used so called the basic model, and it just explained the principle. 
In order to achieve sufficient quantitative security, the randomization technique described here is necessary.
In the above criteria, Y-00 scheme has a potential to improve quantitative security by additional randomization technology,
 because all physical parameters are finite.
In this point of view, we have developed a new concept such as ``quantum noise diffusion technology"[18,19].  
In addition, several randomizations based on Yuen's idea [6]  have been discussed [20].

{\it ``Although we have, at present, no general theory on randomization, using these techniques, it is expected to have security performance that cannot be achieved by conventional cipher. One of them is a special relation between secret key and data (plaintext). That is, under the condition of $H(X_n \mid C_n^B, K_s)=0$, 
one can expect the following security performance"}:
\begin{equation}
H(X_n \mid C_n^E, K_s) \ne 0
\end{equation}
for certain finite $n > |K_s|$. 
$n$ is the length of the plaintext, 
$C_n^B$ means the ciphertext for Bob (signal received by Bob), and $C_n^E$ means the ciphertext for Eve (signal received by Eve).
This is an amazing capability, and one that cannot be achieved even with any conventional cipher including OTP (see Appendix [C]).
In this way, we can say that Y-00 quantum stream cipher has abillity to provide security that exceeds the performance of 
conventional cryptography while maintaining the capabilities of ordinary optical communication.

\section{ Concrete Applications of Quantum Stream Cipher}
As mentioned above, Y-00 quantum stream ciphers has not yet reached their ideal performance, 
but in practical use, they have achieved a high level of security that cannot be achieved with conventional techniques, 
and it can be said that they are now at a level where they can be introduced to the market. 

Since quantum stream cypher is a physical cypher, it requires a dedicated transmitter and receiver. So far, principle models for commercial purposes have been developed at Northwestern University, Tamagawa University, Panasonic, and Hitachi, Ltd.  Fig.7 shows the transceivers of each research institute. The communication speed is 1 Gbit/sec to 10 Gbit/sec and the communication distance is 100 km to 1,000 km. Using these transceivers, operational tests were conducted in a real optical communication network.
Here, we introduce examples of the use case of  Y-00 quantum stream cipher.\\

\subsection{Optical Fiber Communication}
Large amounts of important data are instantaneously exchanged on the communication lines between data centers 
that various data is accumulated.
 It is important from the viewpoint of system protection to eliminate the risk that the data is copied in its entirety
 from communication channel. 
 We believe that Y-00 quantum stream cipher is the best technology for this purpose.
On the other hand, this technology can be used for optical amplifier relay system.  Hence, it can apply to 
the current optical communication systems. 
Transceivers capable of cryptographic transmission at speeds from one Gbit/sec to 10 Gbit/sec have already been realized, 
and by wavelength division multiplexing, 100 Gbit/sec system has been tested. Also, communication distances of 1,000 km to 
10,000 km have been demonstrated. In off-line experiments, 10 Tbit/sec has been demonstrated.
In general, a dedicated line such as dark fiber is required. 
If we want to apply this technology to network function, 
we need the optical switching technology developed by the National Institute of Advanced Industrial Science and Technology (AIST).  
Thus, in collaboration with AIST and other organizations, we have successfully demonstrated the feasibility of using
Y-00 transceiver in testbed optical switching systems.
Furthermore, the references [21-28] show the recent activities of the experimental research group at Tamagawa University towards
 practical application to the real world.

\begin{figure}
\centering{\includegraphics[width=8cm]{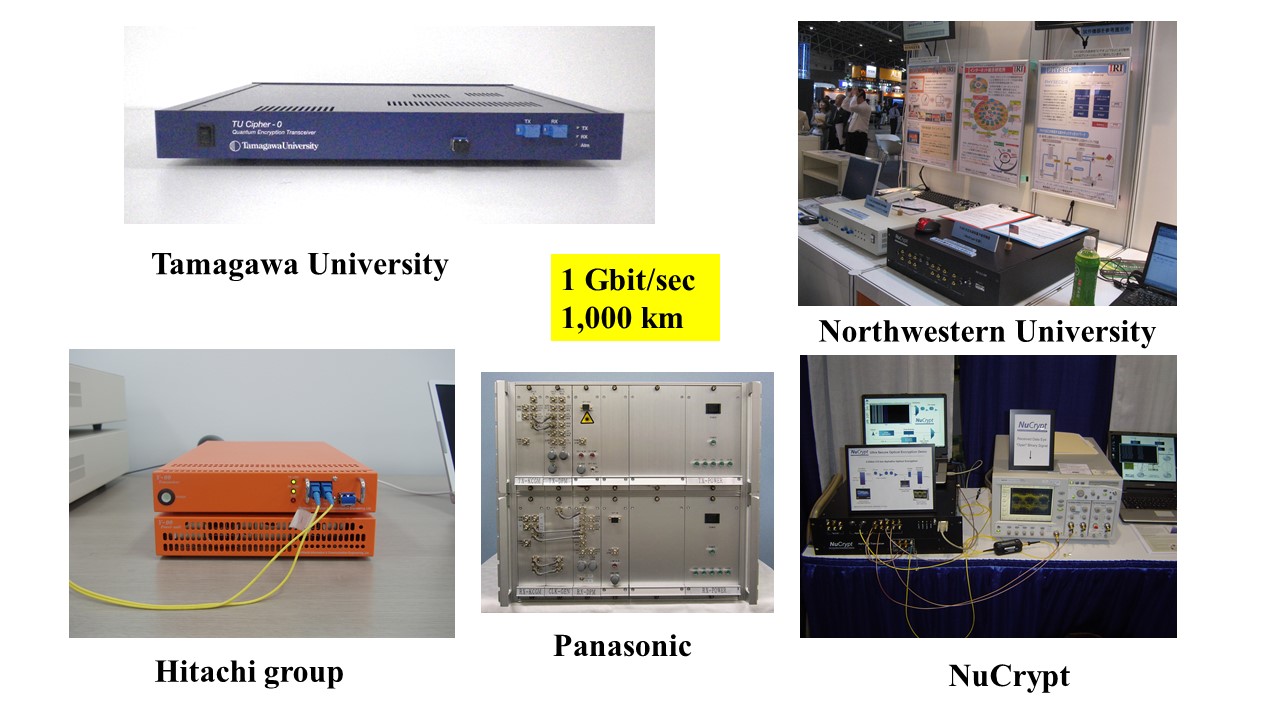}}
\caption{ Commercial Y-00 Transceiver  for 1 Git/sec optical Ethernet. This can be mass produced.}
\end{figure}

\subsection{ Optical Satellite Communication}
Y-00 quantum stream cipher, which was developed for fiber-optic communications, can also be applied to satellite communications.
In satellite communication applications,  rate of operation is an important factor because communication performance depends on weather conditions.
With QKD, it is difficult to keep communications up and running except on clear-air nights.
In the case of Y-00, communication by any satellite system can be almost ensured when the weather is clear.
In case of bad weather, the effects of atmospheric turbulence and scattering phenomena need to be considered.
 We are currently analyzing the performance of the system in such cases at  10 Gbps operation [29].

\subsection{Optical Communication from Base at Moon to Earth}
The Japanese government has initiated a study to increase the user transmission rate of optical space communications from 1.8 Gbps to 
more than 10 Gbps. Furthermore, in the future, the government aims to achieve higher transmission rates in ultra-long distance 
communications required for lunar and planetary exploration. This plan is called LUCAS.
We have started to design for an implementation of 1 Gbps communication system at a transmission distance of 380,000 km
 between the Moon and the Earth using the high-speed performance of the Y-00 quantum stream cipher.

\section{Conclusion}
The current optical network was not laid out in a planned manner, but was configured by extending the existing communication
 lines for adapting the demand.
 In the future, the configuration and specifications of the optical network will be determined following to new urban planning. 
An actual example is the  Smart City that Toyota Motor Corporation et al have disclosed as a future plan. 
Many ideas are also being discussed in other organizations. Recently, NTT has announced a future network concept so called IOWN. 
 In these systems, the security of the all optical network with ultra-high speed is also important issue. 
The group of QKD and the group of Y-00  are promoting their respective technologies.
Y-00 quantum stream cipher is a technology that can realize the specification of high speed and long communication distance. 
In addition, the signals of Y-00 cipher with ultra-multiple valued scheme for coherent state signal, so called quantum mdulation, can have stronger
 quantum properties than QKD in the sense of quantum detection theory.
So, the security is protected by many quantum no-go theorems.
 Although it is difficult to make an accurate prediction, there is a good chance that such a new technology will be used in the future. 
In view of the situation described in this paper,  Y-00 quantum stream cipher will contribute to real world application of 
quantum technology for Society 5.0, and new business development can be expected.
Finally, we would like to introduce that Chinese research institutes have recently been actively working on Y-00 quantum stream cipher.
Fig.8 shows a list of  academic papers on their activities [30-37]. 
It is expected that many research institutes will participate in this technological development.

\begin{figure}
\centering{\includegraphics[width=8cm]{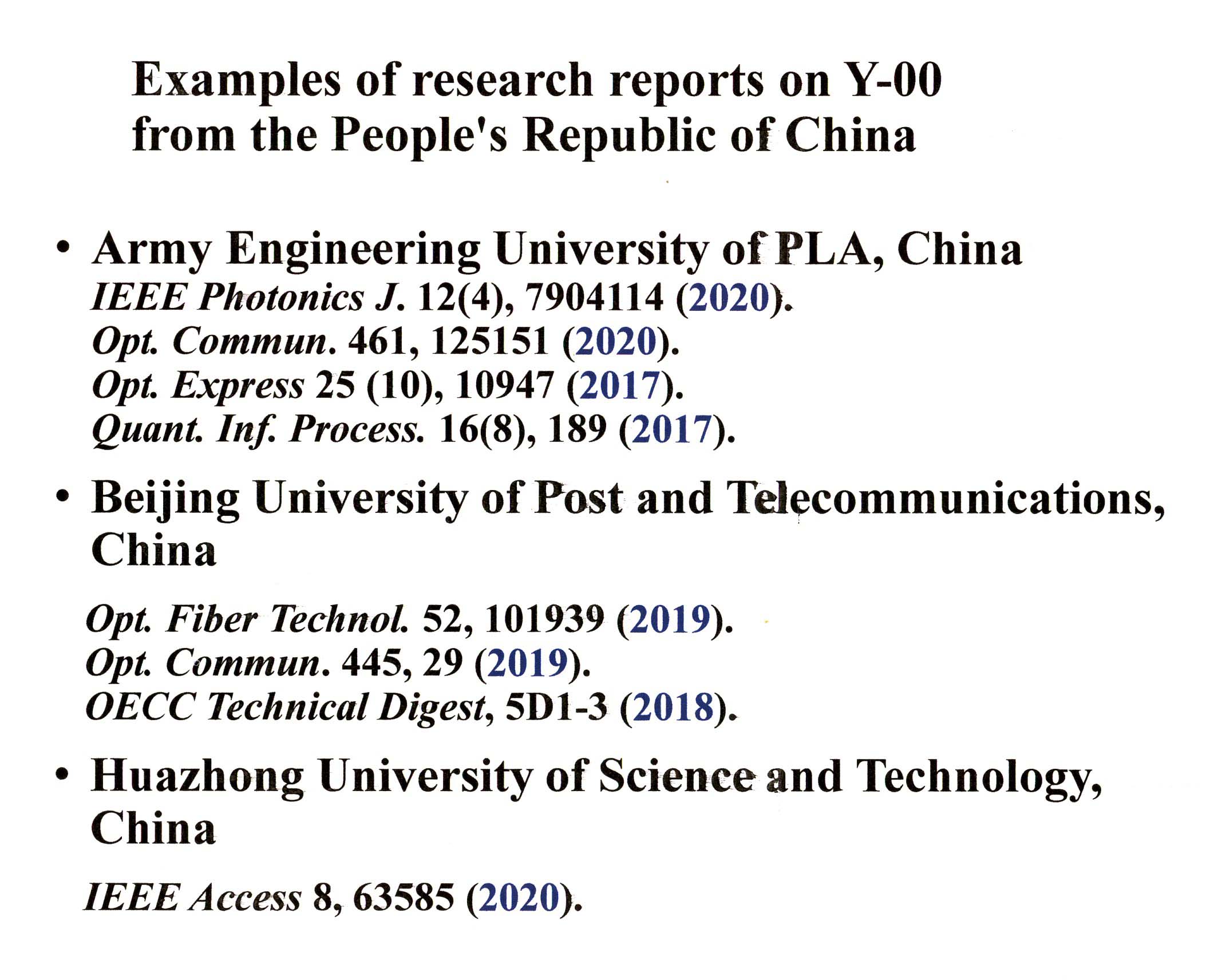}}
\caption{Research activities on Y-00 quantum stream cipher in China }
\end{figure}


\section*{\textbf {Explanation of Symbols}}
Here we give the explanation on the several symbols.\\
\\
(a) Conventional cipher: \\
$X$ is plaintext; $\{0,1\}$, $K_s$ is secret key, $f(K_s) $is running key ;$\{0,1\}$, 
$C$ is conventional ciphertext; $\{0,1\}$.\\
(b) Y-00 quantum stream cipher: \\
 $X$ is plaintext; $\{0,1\}$, $K_s$ is secret key, $f(K_s)$ is running key by PRNG ;$\{0,1\}$,
Y-00 running key is $f(K_s) \mapsto $  $ f_g(K_s)$;  $\{1,2,3,\dots M\}$,
Y-00 ciphertext in the circuit is $y= \alpha_i(X,f_g(K_s),R_p) $;  $\{1,2,3,\dots 2M\}$, 
Y-00 signal is $ \mid \alpha_i(X,f_g(K_s),R_p) >$ =Quantum ciphertext, $C_n^B$ is ciphertext received by Bob:$\{0,1\}$, 
 $C_n^E$ is ciphertext :$\{0,1\}$for Eve transformed from $M$-ary received signal,
$R_p$ is additional randomization.

\appendix

\section* {[A] Quantum computer and quantum computer-resistant cryptography}

It is difficult to predict the realization of a quantum computer capable of cryptanalysis.
 It has been discovered in our recent paper [38] that a new type of error so called nonlinear error or bust error occurs in general quantum computer.
Therein, an error probability for single qubit increases depending on number of qubits in the system. 
These nonlinear errors and bust errors are caused by the recurrence effect due to quantum correlation or the collective decoherence, and by cosmic ray.
 They give a serious damage to scalable quantum computer, and give serious degradations of the capability of quantum computer.
In addition, a number of previously unknown and extremely difficult problems in the development for an error correctable quantum
 computer have been reported.
Thus, the capability of a real quantum computer is strictly limited and that the current cryptography is not subject to the danger posed
 by current quantum computers.
However, we believe that the ideal quantum computer will be realized in the future. 
So, one should develop quantum computer-resistant cryptosystems based on mathematical analysis, or by physical cipher on 
the assumption that an ideal quantum computer or new mathematical discovery can be realized in the future.

Recently, J. P. Mattsson, B. Smeets, and E. Thormarker [39] have provided an excellent  survey for  the NIST quantum computer-resistant
 cryptography standardization effort, the migration to quantum-resistant public-key cryptography, and  the relevance of QKD as a complement to conventional cryptography.
In particular, these algorithms of quantum-resistant public-key cryptography can execute completely in software on classical computers, 
 in contrast to e.g., QKD  which requires very expensive custom hardware. 
For functions of  authentication, signature, and key distribution, such capability provided by software is the most important
 in the real world application.

\section* {[B] Position of security system based on QKD in practical applications}
 To complete  full quantum secure communication systems, at present, we are challenged  to cope with the following problems, discussing a new type of QKD.

Recently, NSA [40], NCSC [41] and ANSSI [42] announced the international stance on QKD. 
They have a negative view of QKD, because the communication performance of QKD based on weak signal is not enough for
 applications to the real situation. Let us denote their comments in the following, respectively.

\subsection{NSA (USA) }
(a)	NSA does not recommend the usage of QKD  for securing the transmission of data in National Security Systems :(NSS) \\
(b)	QKD utilizes the unique properties of quantum mechanical systems to generate and distribute cryptographic keying material
 using special purpose technology. Quantum cryptography uses the same physics principles and similar technology to communicate over
 a dedicated communications link. Published theories suggest that physics allows QKD  to detect the presence of an eavesdropper, 
a feature not provided in standard cryptography.\\

QKD and similar quantum cryptography vendors and the media occasionally state bold claims based on theory e.g., that this technology offers guaranteed security based on the laws of physics. Communications needs and security requirements physically conflict in the use of QKD, and the engineering required to balance these fundamental issues has extremely low tolerance for error. Thus, security of QKD is highly implementation dependent rather than assured be laws of physics. \\

${\bf Technical}$ ${\bf  limitations}$\\
(1)	Quantum key distribution is only a partial solution. \\
(2)	Quantum key distribution requires special purpose equipment.\\
(3)	Quantum key distribution increases infrastructure costs and insider threat risks. \\
(4)	Securing and validating quantum key distribution is a significant challenge.\\
(5)	Quantum key distribution increases the risk of denial of service. \\

For all of these reasons, NSA does not support the usage of QKD to protect communications in National Security Systems, and does not anticipate certifying or approving any QKD security products for usage by NSS customers unless these limitations are overcome.

\subsection{NCSC (UK) }
Given the specialized hardware requirements of QKD over classical cryptographic key agreement mechanisms and the requirement for authentication in all use cases, the NCSC does not endorse the use of QKD for any government or military applications, and cautions against sole reliance on QKD for business critical networks, especially in Critical National Infrastructure sectors. In addition, we advise that any other organizations considering the use of QKD as a key agreement mechanism ensure that robust quantum-safe cryptographic mechanisms for authentication are implemented alongside them. NCSC advice is that the best mitigation against the threat of quantum computers is quantum safe cryptography. Our white paper on quantum-safe cryptography is available on the NCSC website. The NCSC design principles for high assurance systems, which set out the basis under which products and systems should be designed to resist elevated threats, is also available

\subsection{ANSSI (The French National Agency for the Security of Information Systems)  }
Quantum Key Distribution (QKD) presents itself as a technology functionally equivalent to common asymmetric key agreement schemes that are used in nearly all secure communication protocols over the Internet or in private networks. The defining characteristic of QKD is its alleged superior secrecy guarantee that would justify its use for high security applications. However, deployment constraints specific to QKD hinder large-scale deployments with high practical security. Furthermore, new threats on existing cryptography, and in particular the emergence of universal quantum computers, can be countered without resorting to QKD, in a way that ensures the future of secure communications. Although QKD can be used in a variety of niche applications, it is therefore not to be considered as the next step for secure communications
\\

\section* {[C] Drawback of One Time Pad cipher}
OTP is extremely inefficient for the encryption of data, because it requires key sequence as same as data sequence. 
However, it has the following benefit:\\
(1)  Ciphertext only attack on data\\
Since the secret key : $K_s$ is a perfect random number, the ciphertext : $C$ is also a perfect random number. 
Therefore, obtaining the ciphertext gives no information about the plaintext : $X$.  So one has $H(X|C)=H(X)$. At this point, it is called perfect information-theoretic security or unconditional security. \\
(2)  Known plaintext attack on key\\
In OTP, if the length of the known plaintext is $|X|=N$, then the key of the same length $N$ can be known for sure by obtaining a ciphertext $N$ of the same length. However, since the key sequence is completely random, subsequent key sequences cannot be predicted. \\
\\
On the other hand, it has the following drawback:\\
(1)  Falsification attack\\
If Eve obtains the correct ciphertext, and she can invert 0 and 1, and resend. Then 1 and 0 of the data are inverted.
As an example,  if the data is yes or no, the falsification will be successful. So OTP is not secure against falsification attacks [43].\\
(2) Partial known plaintext attack on data\\
If a plaintext sequence has a correlation (e.g., a word), then the possibility of identifying a word arises through a brute force search with
 a partial known plaintext attack [44]. These are some examples of the fact that  OTP does not satisfy the security requirements of modern cryptology.\\

\end{document}